\documentclass[aps,pra,12pt,onecolumn]{revtex4-2}
\usepackage{amsmath}
\usepackage{amssymb}
\usepackage[unicode=true,colorlinks=true,citecolor=blue,urlcolor=blue]{hyperref}
\usepackage{graphicx}

\begin{document}
\title{Braiding of dynamical eigenvalues of Hermitian bosonic Kitaev chains}

\author{Heming Wang}
\affiliation{Department of Electrical Engineering and Edward L. Ginzton Laboratory, Stanford University, Stanford, California 94305, USA}
\author{Shanhui Fan}
\email{shanhui@stanford.edu}
\affiliation{Department of Electrical Engineering and Edward L. Ginzton Laboratory, Stanford University, Stanford, California 94305, USA}

\begin{abstract}
In quantum mechanics, observables correspond to Hermitian operators, and the spectra are restricted to be real.
However, the dynamics of the underlying fields may allow complex eigenvalues and therefore create the possibility of braiding structures.
Here we study the braiding of dynamical eigenvalues in quantum systems by considering Hermitian bosonic Kitaev chains with multiple bands.
The dynamics of the quantum fields in these systems are described by their dynamic matrices, which have complex eigenvalues.
We show that there are symmetry constraints imposed on these dynamic eigenvalues.
Despite these constraints, braiding is possible for frequencies within the effective gain and loss regions of the complex plane.
We explicitly construct two- and three-strand braidings using the exceptional points found in the system and discuss possible implementations.
\end{abstract}
\maketitle

\section{Introduction}
The topological properties of band structures \cite{qi2011topological, lu2014topological, ozawa2019topological} have enriched our understanding of different phases of matter. Traditional topological invariants focus on the eigenvectors (i.e., state compositions) of the system that predict the existence or absence of edge states at the interface, among other observable phenomena. With the introduction of non-Hermiticity in band structures \cite{lee2016anomalous, yao2018edge, yao2018non-hermitian, bergholtz2021exceptional}, the concept of topology has been extended to eigenvalues, which may exhibit winding and braiding \cite{shen2018topological, zhang2020correspondence, hu2021knots, wang2024non-hermitian}. Specifically, the braiding behavior of the eigenvalues provides a classification of non-Hermitian systems \cite{wojcik2020homotopy, li2021homotopical, wojcik2022eigenvalue, hu2022knot} that is different from the eigenvector topology. These are closely related to the existence of exceptional points and non-abelian dynamics \cite{parto2020non-hermitian, guo2023exceptional} and have been demonstrated by experiments \cite{wang2021topological, patil2022measuring, tang2022experimental, zhang2023observation, li2023eigenvalue, cao2023probing, zhang2023experimental, wu2023observation, guria2024resolving, long2024non-abelian}.

Despite the theoretical interest and potential applications of energy braiding, most of the experimental implementations and proposed theoretical models \cite{konig2023braid-protected, zhong2023numerical, zhu2024versatile, rafi-ul-islam2024knots} are focused on classical systems. Here we extend energy band braiding into quantum systems using the Hermitian bosonic Kitaev chain (BKC) as an example. The BKC \cite{mcdonald2018phase-dependent, wan2023quantum-squeezing-induced} is a versatile quantum platform that has already demonstrated squeezing and non-reciprocal quadrature transport. The Hamiltonian describing the BKC satisfies Hermiticity constraints. However, the dynamics of the underlying fields, which are described by the dynamic matrices, may exhibit growth or decay. As a result, the dynamic eigenvalues, i.e., the eigenvalues of the dynamic matrices, may be complex and can lead to braiding structures.

The paper is organized as follows.
In Section \ref{sec:BKC} we introduce the BKC and discuss the existence of complex dynamic eigenvalues.
In Section \ref{sec:EP} we analyze the symmetry constraints on the dynamic eigenvalues and provide examples of exceptional points.
In Section \ref{sec:braid} we explicitly construct models with nontrivial braidings of dynamical eigenvalues.
Finally, in Section \ref{sec:summary} we discuss potential realizations of the models and conclude.

\section{Hermitian bosonic Kitaev chains} \label{sec:BKC}

\begin{figure}
\centering
\includegraphics[width=85mm]{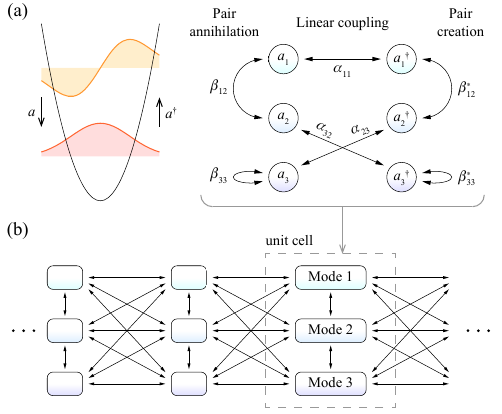}
\caption{Generic Hamiltonians that are quadratic with respect to bosonic operators.
(a) Left: creation and annihilation operators for a harmonic oscillator. Right: Some possible processes that are quadratic in the bosonic operators, using $r=3$ as an example. Each process correspond to a single term in Eq. (\ref{eq:H_unit}).
(b) Using the system in (a) as unit cells and chaining them together creates a bosonic Kitaev chain, where the couplings between unit cells become the Fourier coefficients of the coupling as a function of $K$ in the wavevector space.}
\label{fig:1}
\end{figure}

We consider the following general Hamiltonian that is quadratic with respect to a set of $r$ bosonic operators [Fig. \ref{fig:1}(a)],
\begin{equation}
H_\text{unit} = \sum_{1 \leq j,k \leq r}\alpha_{jk} a_j^\dagger a_k + \sum_{1 \leq j,k \leq r}\left(\frac{\beta^*_{jk}}{2} a_j^\dagger a_k^\dagger + \frac{\beta_{jk}}{2}a_j a_k\right)
\end{equation}
where $j$ and $k$ are the indices of the bosonic modes with $1\leq j,k\leq r$, $a_j^\dagger$ and $a_j$ are the raising and lowering operators for the mode $j$, respectively, $\alpha_{jk}$ are the coupling constants (for $j=k$, $\alpha_{jj}$ is the energy of the mode $j$), and $\beta_{jk}$ are the pair creation and annihilation constants. Hermiticity requires that $\alpha_{jk}^* = \alpha_{kj}$. Although there are no a priori constraints on $\beta$, here it is sufficient to set $\beta_{jk} = \beta_{kj}$ to maintain the symmetric form of $H$.

The Hamiltonian can be more readily analyzed in its quadrature form. To this end, we define the following quadrature operators:
\begin{equation}
x_j = \frac{1}{\sqrt{2}}\left(a_j + a_j^\dagger\right),\ \ p_j = \frac{1}{i\sqrt{2}}\left(a_j - a_j^\dagger\right)
\end{equation}
where $i^2=-1$. The Hamiltonian can be rewritten using the quadrature operators as
\begin{equation}
H_\text{unit} = \sum_{1 \leq j,k \leq r} \frac{t_{jk}}{2} p_jp_k + \sum_{1 \leq j,k \leq r} \frac{v_{jk}}{2} x_jx_k + \sum_{1 \leq j,k \leq r} g_{jk}x_jp_k
\label{eq:H_unit}
\end{equation}
where constant terms arising from the commutators have been dropped, and the coefficients are given by
\begin{align}
t_{jk} &= \frac{1}{2}\left(\alpha_{jk} + \alpha_{kj} - \beta_{jk} - \beta_{jk}^*\right) \\
v_{jk} &= \frac{1}{2}\left(\alpha_{jk} + \alpha_{kj} + \beta_{jk} + \beta_{jk}^*\right) \\
g_{jk} &= \frac{i}{2}\left(\alpha_{jk} - \alpha_{kj} + \beta_{jk} - \beta_{jk}^*\right)
\end{align}
These coefficients satisfy $t_{jk} = t_{kj}$ and $v_{jk} = v_{kj}$. In addition, Hermiticity requires that all $t_{jk}$, $v_{jk}$, and $g_{jk}$ be real. We denote $\psi \equiv (X; P)$ with $X \equiv (x_1, x_2, \cdots x_r)^T$ and $P \equiv (p_1, p_2, \cdots p_r)^T$. In the Heisenberg picture, the equation of motion for $\psi$ can be found as
\begin{equation}
\frac{d}{dt} \begin{bmatrix} X \\ P \end{bmatrix} = 
\begin{bmatrix}
\partial H_\text{unit}/\partial P \\ -\partial H_\text{unit}/\partial X
\end{bmatrix} =
\begin{bmatrix} G^T & T \\ -V & -G \end{bmatrix}
\begin{bmatrix} X \\ P \end{bmatrix}
\label{eq:unit_dynamics}
\end{equation}
with $T$, $V$ and $G$ being the matrices formed by $t_{jk}$, $v_{jk}$ and $g_{jk}$, respectively. As such, the dynamics of the operators $\psi$ is described by its dynamic matrix:
\begin{equation}
H \equiv \begin{bmatrix} G^T & T \\ -V & -G \end{bmatrix}
\label{eq:unit_matrix}
\end{equation}

We next construct a BKC by using the system described by Eq. (\ref{eq:H_unit}) as unit cells and chaining them together. The resulting periodic structure can be described by
\begin{equation}
H_\text{BKC} = \sum_{m,n}\sum_{j,k}\left[\frac{t_{jk,m-n}}{2} p_{j,m}p_{k,n} + \frac{v_{jk,m-n}}{2} x_{j,m} x_{k,n} + g_{jk,m-n}x_{j,m} p_{k,n}\right]
\label{eq:H_BKC_chain}
\end{equation}
where $m$ and $n$ are integers that denote the positions of cells along the chain, and the couplings depend only on the relative distance $m-n$. The periodic chain admits plane-wave states with conserved wavevectors, which can be defined as $\psi(K) = \sum_n \psi_n \exp(-iKn)$ with $K$ the wavevector. The dynamics of $\psi(K)$ is $d\psi(K)/dt = H(K)\psi(K)$ similar to Eq. (\ref{eq:unit_dynamics}), where the dynamic matrix in the wavevector domain can be found as
\begin{align}
H(K) \equiv \begin{bmatrix} G(K)^\dagger & T(K) \\ -V(K) & -G(K) \end{bmatrix}
\label{eq:H_BKC}
\end{align}
with the Fourier transform of the coupling coefficient given by
\begin{equation}
t_{jk}(K) = \sum_n t_{jk,n} \exp(-iKn)
\end{equation}
and similarly for $v_{jk}(K)$ and $g_{jk}(K)$. The Hermiticity constraint becomes $T(K) = T(K)^\dagger = T(-K)^*$, $V(K) = V(K)^\dagger = V(-K)^*$, and $G(K) = G(-K)^*$ for all $K$. A detailed study of $H(K)$ would reveal its physical properties, including the braiding of its eigenvalues.

We now briefly discuss the relation between the eigenvalues of $H$ and the eigenvalues of $H_\text{unit}$. In some cases, $H$ can be diagonalized and all eigenvalues are purely imaginary:
\begin{equation}
U H U^{-1} = \text{diag}(-i\lambda_{r}, -i\lambda_{r-1}, \cdots, -i\lambda_1, i\lambda_1, i\lambda_2, \cdots, i\lambda_r)
\end{equation}
where $U$ is the basis transformation and $\lambda_j > 0$ (the pairing of eigenvalues is required by Hamiltonian symmetry, as discussed in the next section). The transformation $U$ defines pairs of ladder operators that are complex conjugates of each other and therefore provides a set of Bogoliubov transformations that can be applied to $H_\text{unit}$. The supermodes in the Bogoliubov transformations are linear combinations of $a_j$ and $a_j^\dagger$, respect the commutation relations for ladder operators, and convert $H_\text{unit}$ into a sum of number operators. As such, the collection of $\lambda_j$ is contained in the eigenvalues of $H_\text{unit}$ as single-particle energy levels.
However, as will be shown below, it is possible that $H$ could not be diagonalized or its eigenvalues are not purely imaginary, which would represent effective gains and losses. The existence of gains and losses for $H$ seems to violate the spectral theorem that self-adjoint operators can only possess a real spectrum. The resolution relies on the fact that the eigenvalues of $H$ and the spectrum of $H_\text{unit}$ are no longer related when $H_\text{unit}$ is not bounded from below and above. For example, with a single bosonic mode ($r=1$), the parameters $g_{11} = 0$, $t_{11} = 1$ and $v_{11} = -1$ in Eq. (\ref{eq:H_unit}) lead to the ``inverted harmonic oscillator'' Hamiltonian $(p^2-x^2)/2$. This Hamiltonian is self-adjoint according to Nelson's commutator theorem \cite{nelson1972time-ordered, faris1974commutators}, which is a result of the photon numbers growing at most exponentially. However, since a ground state does not exist, the usual ladder operator method to construct the spectrum is no longer applicable \cite{he2025hidden}. This discussion extends to $H(K)$, which may also possess generic complex eigenvalues despite being derived from the Hermitian Hamiltonian $H_\text{BKC}$. In the following, we focus on the eigenvalues of $H(K)$ as a matrix and their braiding properties.

\section{Exceptional points} \label{sec:EP}

\begin{figure}
\centering
\includegraphics[width=85mm]{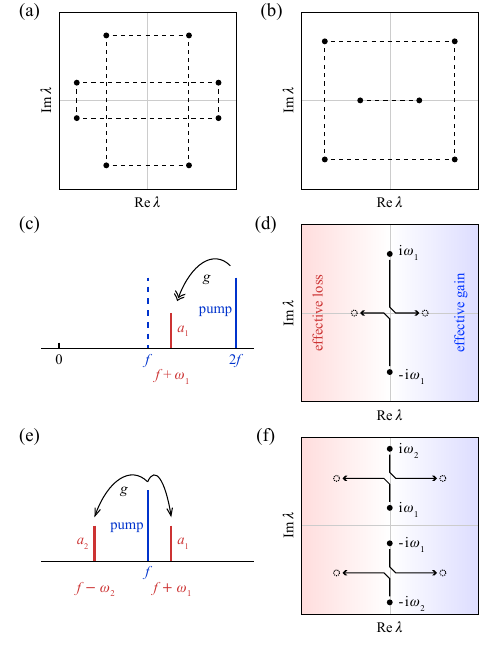}
\caption{Symmetries of eigenvalues and exceptional points in Hamiltonian matrices.
(a) An example of a model with $r=4$, where each quadrant contains two eigenvalues, and the eight eigenvalues are symmetric with respect to reflections across the axes.
(b) An example of a model with $r=3$, where each quadrant contains one eigenvalue, and the other two eigenvalues are on the real axis.
(c) The model Eq. (\ref{eq:ex_1}) as an optical parametric oscillator.
(d) The eigenvalue trajectory of Eq. (\ref{eq:ex_1}) as $g$ increases from $0$. An exceptional point can be found at $\lambda = 0$.
(e) The model Eq. (\ref{eq:ex_2}) as a four-wave mixing process.
(f) The eigenvalue trajectory of Eq. (\ref{eq:ex_2}) as $g$ increases from $0$. Two exceptional points can be found at $\lambda = \pm i(\omega_1+\omega_2)/2$.
}
\label{fig:2}
\end{figure}

In this section, we study the possible eigenvalues of $H(K)$ and the restrictions arising from the Hermiticity constraints on $H_\text{unit}$ and $H_\text{BKC}$. In particular, $H(K)$ may not be diagonalizable and contains exceptional points. These are important in constructing eigenvalue braidings, as braid transitions can be induced when a loop in the parameter space crosses the exceptional points.

We first consider the time-reversal symmetric wavevectors $K=0$ and $K=\pi$, hereafter collectively referred to as $\sin K=0$. At these points, $H(K)$ is real and has the same symmetry restrictions as in Eq. (\ref{eq:unit_matrix}), which can be succinctly expressed as:
\begin{equation}
JHJ = H^T,\ \ J=i\sigma_2 \otimes I_r = \begin{bmatrix}  & I_r \\ -I_r &  \end{bmatrix}
\label{eq:ham_matrix_def}
\end{equation}
where $\sigma_2 = [0,-i;i,0]$ is the second Pauli matrix and $\otimes$ is the Kronecker matrix product. Equation (\ref{eq:ham_matrix_def}) is equivalent to the statement that $JH = (JH)^T$. A matrix with real entries satisfying Eq. (\ref{eq:ham_matrix_def}) is known as a Hamiltonian matrix \cite{meyer1992introduction}. The distributions of eigenvalues of Hamiltonian matrices are highly symmetrical: If $\lambda$ is an eigenvalue of $H$, then $\lambda^*$ is an eigenvalue (since $H$ is real), $-\lambda$ is an eigenvalue (since $J^{-1}HJ = -H^T$) and $-\lambda^*$ is also an eigenvalue [Fig. \ref{fig:2}(a)]. This implies that at least one pair of eigenvalues is confined to the real or imaginary axis when $r$ is odd [Fig. \ref{fig:2}(b)].

Unlike Hermitian matrices, Hamiltonian matrices are not always diagonalizable. We demonstrate this using the most general model of a unit cell with two dynamical modes ($r=1$). The Hamiltonian is given by
\begin{equation}
H_\text{unit} = \omega_1 a_1^\dagger a_1- \frac{g}{2}\left[\left(a_1^\dagger\right)^2+a_1^2\right]
\label{eq:ex_1}
\end{equation}
with $\alpha_{11} = \omega_1$ and $\beta_{11} = -g$. The pair generation constant $-g$ is taken to be real, as any complex phase of $\beta$ can be absorbed into the definition of $a_1$ and $a_1^\dagger$. Equation (\ref{eq:ex_1}) could represent a single-mode optical parametric oscillator with phase mismatch [Fig. \ref{fig:2}(c)]. Its dynamical matrix is given by
\begin{equation}
H = \begin{bmatrix}
&\omega_1+g \\ -\omega_1+g &
\end{bmatrix}
\end{equation}
The eigenvalues of $H$ are given by $\lambda = \pm i\sqrt{\omega_1^2-g^2}$. These eigenvalues are purely imaginary when $g<\omega_1$, where the ground state of $H_\text{unit}$ is squeezed with respect to $a_1$. When $g>\omega_1$, the eigenvalues become real, and the amplification process occurs. The transition occurs at $g=\omega_1$, which is an exceptional point in the system, corresponding to $\lambda = 0$ [Fig. \ref{fig:2}(d)].

A more complicated example with four dynamical modes ($r=2$) is given by
\begin{equation}
G = \mathbf{0}_{2\times2},\ \ 
T = \begin{bmatrix} \omega_1 & g \\ g & -\omega_2 \end{bmatrix},\ \ 
V = \begin{bmatrix} \omega_1 & -g \\ -g & -\omega_2 \end{bmatrix}
\end{equation}
\begin{equation}
H_\text{unit} = \omega_1 a_1^\dagger a_1 - \omega_2 a_2^\dagger a_2 - g\left(a_1^\dagger a_2^\dagger + a_1a_2\right)
\label{eq:ex_2}
\end{equation}
which can be realized by a four-wave mixing process with two sidebands, as seen in primary comb generations in optical frequency combs \cite{chembo2010spectrum} [Fig. \ref{fig:2}(e)]. The eigenvalues can be found as $\lambda = i [\pm(\omega_1+\omega_2)\pm\sqrt{(\omega_1-\omega_2)^2-4g^2}]/2$. Exceptional points occur at $\lambda = \pm i(\omega_1+\omega_2)/2$ when $g = |\omega_1-\omega_2|/2$ [Fig. \ref{fig:2}(f)], corresponding to the primary comb generation threshold.

The appearance of exceptional points in $H$ is a general feature, and their locations are not necessarily restricted to the real or imaginary $\lambda$ axes. To demonstrate this, we consider the following model with eight dynamical modes ($r=4$):
\begin{equation}
G = \mathbf{0}_{4\times 4}
\label{eq:EP_08G}
\end{equation}
\begin{equation}
T = \begin{bmatrix}
  &   & 1 & 1 \\
  &   & 1 &-1 \\
 1& 1 &   &   \\
 1&-1 &   &   
\end{bmatrix}
\label{eq:EP_08T}
\end{equation}
\begin{equation}
V = \begin{bmatrix}
\delta &   & 1 &-1 \\
  & -\delta &-1 &-1 \\
 1&-1 &   &   \\
-1&-1 &   &  
\end{bmatrix}
\label{eq:EP_08V}
\end{equation}
When $\delta = 0$, the dynamic matrix splits into two blocks, with one block spanned by $x_1,x_2,p_3,p_4$ and the other block by $p_1,p_2,x_3,x_4$. Each block has a structure identical to that in the four-mode example above. As a result, the model features four pairs of degenerate modes at $\lambda = \pm 1 \pm i$. The introduction of a nonzero $\delta$ does not change the eigenvalues but destroys one eigenmode at each eigenvalue, leading to an exceptional point in each quadrant.

\begin{figure}
\centering
\includegraphics[width=85mm]{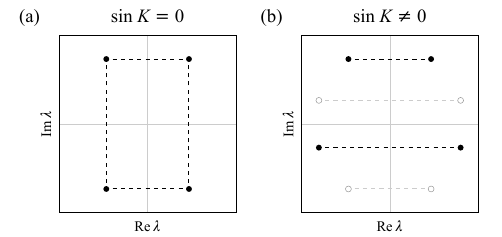}
\caption{Comparisons of eigenvalue distributions of $H(K)$.
(a) At $\sin K = 0$, the eigenvalues possess reflection symmetry across both real and imaginary axes.
(b) At $\sin K \neq 0$, only the reflection symmetry across the imaginary axis is guaranteed. Reflection across the real axis transforms the eigenvalues of $H(K)$ (solid circles) to eigenvalues of $H(-K)$ (empty circles).
}
\label{fig:3}
\end{figure}

We now consider the general case of $H(K)$ when $\sin K\neq 0$. The entries of $H(K)$ need not be real, and the matrix $H(K)$ satisfies the following:
\begin{equation}
JH(K)J = H(K)^\dagger,\ \ J = \begin{bmatrix}  & I_r \\ -I_r &  \end{bmatrix}
\end{equation}
This indicates that each $H(K)$ matrix with $\sin K \neq 0$ has fewer symmetry constraints than $H$. If $\lambda$ is an eigenvalue of $H(K)$, then only $-\lambda^*$ is guaranteed to be another eigenvalue (using $J^{-1}H(K)J = -H(K)^\dagger$). As such, the eigenvalues are symmetric across the imaginary axes but not necessarily across the real axis (Fig. \ref{fig:3}). This allows for more configurations of eigenvalues and exceptional points, such as a single pair of exceptional points off the real axis. However, the reflections of the eigenvalues of $H(K)$ across the real axis are the eigenvalues of $H(-K)$ since $H(K)=H(-K)^*$.

\section{Eigenvalue braiding} \label{sec:braid}

The existence of exceptional points in $H$ and $H(K)$ as discussed in the previous section can be used to provide examples of braiding of dynamical eigenvalues. The $2r$ eigenvalues of $H(K)$ form strands in the $(\text{Re}\ \lambda, \text{Im}\ \lambda, K)$ space. In the case of a constant $H(K)$, the eigenvalues do not move in the $\lambda$ plane, and the strands form trivial braids. If the trajectory of $H(K)$ is deformed and passes through an exceptional point, the strands will intersect each other and the connections between the endpoints will change. As such, explicit examples of braidings can be constructed by starting from $H$ that features exceptional points and applying perturbations.

The eigenvalues of $H(K)$ possess mirror symmetry with respect to the imaginary axis in the complex $\lambda$ plane. When an eigenvalue on the left half-plane moves toward the imaginary axis, it will collide with another eigenvalue approaching the axis from the other side. Therefore, for the potential braiding strands to not intersect each other, each strand should be contained within the left (effective loss) or right (effective gain) half-plane. A minimum of four dynamical modes ($r \geq 2$) are necessary for at least two dynamical eigenvalues on each half-plane and form nontrivial braids. The braids also satisfy the mirror symmetry across the imaginary axis, and the braiding of all dynamical eigenvalues is restricted by this symmetry. In addition, since $H(K)=H(-K)^*$, the braiding must also possess a ``time-reversal symmetry'' defined by $\lambda \rightarrow\lambda^*$ and $K\rightarrow -K$. This may be satisfied by a single group of strands that intersects the real axis or two groups of strands separated by the real axis.

\begin{figure}
\centering
\includegraphics[width=85mm]{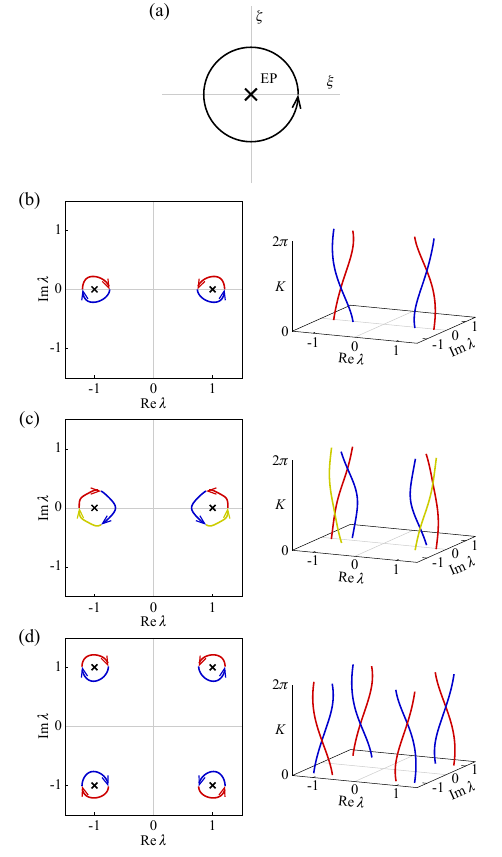}
\caption{Dynamical eigenvalue braiding in bosonic Kitaev chains.
(a) The trajectory $\xi = \rho \cos K$, $\zeta = \rho \sin K$ in the $(\xi,\zeta)$ parameter space with respect to $K$ circles around an exceptional point at $\xi=\zeta=0$.
(b)(c)(d) Dynamical eigenvalue trajectories for Eq. (\ref{eq:braid_04}), Eq. (\ref{eq:braid_06}), and Eqs. (\ref{eq:braid_08G}-\ref{eq:braid_08V}), respectively. Here $\xi = \rho \cos K$, $\zeta = \rho \sin K$, $\rho=0.2$. Left panels show the projection of the strands on the $\lambda$ plane together with the locations of the exceptional points (crosses). Right panels show the braiding of the eigenvalue strands. Colors are for differentiating the strands only.
}
\label{fig:4}
\end{figure}

We now construct examples of two- and three-strand braidings subject to the symmetry constraints discussed above. We consider the following model with four dynamical modes ($r=2$):
\begin{equation}
G = \begin{bmatrix}
  & 0 \\
-i\zeta/2 & 
\end{bmatrix},\ 
T = \begin{bmatrix}
  & -1 \\
-1& 
\end{bmatrix},\ 
V = \begin{bmatrix}
1 & 1 \\
1 & \xi
\end{bmatrix}
\label{eq:braid_04}
\end{equation}
At $\xi = \zeta = 0$, the model features a pair of exceptional points at $\lambda = \pm 1$. For small values of $\xi$ and $\zeta$, the dynamical eigenvalues near the exceptional point $1$ can be found as, up to the lowest order,
\begin{equation}
\lambda \approx 1 \pm \frac{1}{2}\sqrt{\xi + i\zeta}
\end{equation}
The dynamical eigenvalues near $\lambda = -1$ can be obtained by symmetry. As a concrete example, we can choose $\xi = \rho \cos K$ and $\zeta = \rho \sin K$ with $\rho \ll 1$. The choice of $\xi$ and $\zeta$ satisfies $\xi(K) = \xi(-K)$ and $\zeta(K) = -\zeta(-K)$, and is consistent with the Hermiticity constraints. The trajectory of $\xi$ and $\zeta$ with respect to $K$ circles around the exceptional point $\xi=\zeta=0$ in the parameter space [Fig. \ref{fig:4}(a)]. Consequently, the two strands close to $\lambda = 1$ satisfy $\lambda \approx 1\pm \sqrt{\rho}e^{iK/2}/2$. Each strand circles around the point $\lambda = 1$ for half a cycle and is connected to the other strand when $K$ varies from $0$ to $2\pi$. This creates a single braid crossing on the right side of the $\lambda$ plane [Fig. \ref{fig:4}(b)]. We note that the braiding for all dynamical eigenvalues consists of two braid crossings due to symmetry, with a single crossing on each side of the imaginary axis with opposite orientations.

For braids that involve more strands, the braid structure can be constructed by concatenating individual braid crossings, encircling high-order exceptional points, or a combination of both. We demonstrate encircling a third-order exceptional point by a model with six dynamical modes ($r=3$):
\begin{equation}
G = \begin{bmatrix}
  &   & 0 \\
  & 0 &   \\
-i\zeta/2 & &
\end{bmatrix},\ 
T = \begin{bmatrix}
  &    & -1 \\
  & -1 & \\
-1& &
\end{bmatrix},\ 
V = \begin{bmatrix}
  & 1 & 1\\
1 & 1 &  \\
1 &   &\xi
\end{bmatrix}
\label{eq:braid_06}
\end{equation}
The model described by Eq. (\ref{eq:braid_06}) has a structure similar to that of Eq. (\ref{eq:braid_04}) and features a pair of third-order exceptional points, as seen in the expansion $\lambda \approx 1 + (\xi+i\zeta)^{1/3}/2$. For $\xi = \rho \cos K$ and $\zeta = \rho \sin K$ with $\rho \ll 1$, the dynamical eigenvalues on the right half-plane become $\lambda \approx 1+e^{2\pi ik/3}\rho^{1/3}e^{iK/3}/2$ with $k=0,1,2$. Each strand circles around the exceptional point $\lambda = 1$ by $2\pi/3$ and connects to the next strand in the counterclockwise direction [Fig. \ref{fig:4}(c)].

It is also possible to create braidings that do not intersect the real $\lambda$ axis. This can be achieved by perturbing the model with eight dynamical modes ($r=4$) as described by Eqs. (\ref{eq:EP_08G}-\ref{eq:EP_08V}):
\begin{equation}
G = \begin{bmatrix}
 & & & 0 \\
 & & 0 & \\
 &-i\zeta/2 & & \\
 i\zeta/2 & & &
\end{bmatrix}
\label{eq:braid_08G}
\end{equation}
\begin{equation}
T = \begin{bmatrix}
  &   & 1 & 1 \\
  &   & 1 &-1 \\
 1& 1 &   &   \\
 1&-1 &   &   
\end{bmatrix}
\label{eq:braid_08T}
\end{equation}
\begin{equation}
V = \begin{bmatrix}
1 &   & 1 &-1 \\
  &-1 &-1 &-1 \\
 1&-1 & -\xi & \\
-1&-1 & & \xi
\end{bmatrix}
\label{eq:braid_08V}
\end{equation}
The dynamical eigenvalues in the first quadrant can be found as $\lambda \approx 1+i\pm\sqrt{\xi+i\zeta}/2$. The choice $\xi = \rho \cos K$, $\zeta = \rho \sin K$, and $\rho \ll 1$ produces the desired braiding patterns [Fig. \ref{fig:4}(d)]. The braiding for all dynamcial eigenvalues consists of four braid crossings due to symmetry, with a single braid crossing in each quadrant. Two crossings across the imaginary axis have opposite orientations, but two crossings across the real axis have the same orientation due to the ``time-reversal symmetry'' defined by $\lambda \rightarrow\lambda^*$ and $K\rightarrow -K$. Finally, we note that braidings that encircle higher-order exceptional points within individual quadrants can be generated similarly by expanding the dimensions of the $G$, $T$ and $V$ matrices. 

\section{Final remarks and conclusion} \label{sec:summary}

As a final remark, we now briefly comment on the realization of $H_\text{BKC}$. Models with a few bosonic modes \cite{slim2024optomechanical, busnaina2024quantum, lustig2025quadrature-dependent} have been experimentally demonstrated. For additional modes, each $a_j$ mode can be realized as a photonic state in photonic systems. The $\alpha_{jk}$ couplings can then be realized as linear couplings between these states. On the other hand, the $\beta_{jk}$ couplings can be realized by second-order nonlinear processes such as difference frequency generation. This requires the introduction of pump light into the system. Each pump can be mode-matched to selectively generate photon pairs in designated modes to implement each term of $\beta_{jk}$. The pump dynamics can be neglected if we use the undepleted-pump assumption and work in the frequency reference frame that rotates with the pump light. This would realize $H_\text{BKC}$ as an effective Hamiltonian, and the existence of pump provides the energy necessary for the exponential growth of modes.

In conclusion, we have explicitly demonstrated dynamic eigenvalue braiding in bosonic Kitaev chains, thereby extending the concept of braiding into the quantum regime. Unlike the spectrum of Hermitian operators, the dynamic matrix may possess complex eigenvalues, which are necessary for braid formations. Explicit examples of braiding have been constructed by utilizing exceptional points found in Hamiltonian matrices. Although the proposed system requires channels that exchange energy with the outside, the quantum behavior of the bosonic modes involved remains intact, and quantum effects such as pair generations and squeezing can be readily achieved. As such, we believe that the current results will motivate further studies on the quantum effects of dynamical eigenvalue braiding.

This work is funded by a Simons Investigator in Physics grant from the Simons Foundation (Grant No. 827065).

\end{document}